\documentclass[
twocolappendix,
twocolumn,
trackchanges
]{aastex631}

\usepackage{CJKutf8}
\usepackage{hyperref}
\usepackage{amsmath}
\usepackage{physics}
\usepackage{amssymb}
\usepackage{enumerate}
\usepackage{bm}
\usepackage{xspace}
\usepackage{times}

\newcommand{\eg}{e.g.,~}
\newcommand{\ie}{i.e.,~}
\newcommand{\npem}{$npe$ matter\xspace}
\newcommand{\npemm}{$npe\mu$ matter\xspace}

\newcommand{\orcid}[1]{\href{https://orcid.org/#1}{
    \includegraphics[width=10pt]{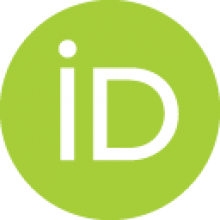}}}

\newcommand{\M}{\texttt{Margherita}~}

\begin{document}

\title{Accurate muonic interactions in neutron star mergers and impact on
  heavy-element nucleosynthesis }

\author{Harry Ho-Yin Ng\:\orcid{0000-0003-3453-7394}}
\affiliation{Institut f\"ur Theoretische Physik, Goethe Universit\"at,
  Max-von-Laue-Str. 1, 60438 Frankfurt am Main, Germany}

\author{Carlo Musolino\:\orcid{0000-0002-9955-3451}}
\affiliation{Institut f\"ur Theoretische Physik, Goethe Universit\"at,
  Max-von-Laue-Str. 1, 60438 Frankfurt am Main, Germany}

\author{Samuel D. Tootle\:\orcid{0000-0001-9781-0496}}
\affiliation{Department of Physics, University of Idaho, Moscow, ID 83844, USA}

\author{Luciano Rezzolla\:\orcid{0000-0002-1330-7103}}
\affiliation{Institut f\"ur Theoretische Physik, Goethe Universit\"at,
  Max-von-Laue-Str. 1, 60438 Frankfurt am Main, Germany}
\affiliation{School of Mathematics, Trinity College, Dublin 2, Ireland}
\affiliation{Frankfurt Institute for Advanced Studies,
  Ruth-Moufang-Str. 1, 60438 Frankfurt am Main, Germany}

\date{\today}

\begin{abstract}
The abundances resulting from $r$-process nucleosynthesis as predicted by
simulations of binary neutron star (BNS) mergers remain an open question
as the current state of the art is still restricted to three-species
neutrino transport. We present the first BNS merger simulations employing
a moment-based general-relativistic neutrino transport with five neutrino
species, thus including (anti)muons and advanced muonic
$\beta$-processes, and contrast them with traditional three-neutrino-species 
simulations. Our results show that a muonic
trapped-neutrino equilibrium is established, forming a different
trapped-neutrino hierarchy akin to the electronic equilibrium. The
formation of (anti)muons and the muonization via muonic $\beta$-processes
enhance neutrino luminosity, leading to {a stronger} cooling in the
early postmerger phase. Since muonic processes redirect part of the
energy otherwise used for protonization by electronic processes, they
yield a cooler remnant and disk, together with neutrino-driven winds that
are more neutron-rich. Importantly, the unbound ejected mass is smaller
than in three-species simulations, and, because of its comparatively
smaller temperature and proton fraction, it can enhance lanthanide
production and reduce the overproduction of light $r$-process elements
for softer equations of state. This finding underlines the importance of
muonic interactions and five neutrino species in long-lived BNS remnants.
\end{abstract}


\section{Introduction}

The ejection of matter from binary neutron star (BNS) mergers is one of
the best sources for the synthesis of heavy elements via the $r$-process
(see~\citet{Baiotti2016, Paschalidis2016, Metzger2017, Arcones2023} for
some reviews). The amount of this ejecta produced by shocks,
neutrino-driven or magnetically driven winds, has been predicted by
numerical simulations (see, \eg \citet{Rosswog1999, Dessart2009,
  Rezzolla:2010, Roberts2011, Kyutoku2012, Rosswog2013a, Bauswein2013b,
  Hotokezaka2013, Foucart2014, Perego2014, Just2015, Martin2015,
  Radice2016, Lehner2016, Dietrich2016, Fujibayashi2017, Bovard2017,
  Kawaguchi2018b, Bernuzzi2024}) and has found support in the GW170817
event (see, \eg \citet{Kasen2017, Rosswog2017b}).

The kilonova signal associated with AT 2017gfo displayed both ``blue'' and
``red'' components, attributed to lanthanide-poor, low-opacity ejecta
(with electron fraction $Y_e \gtrsim 0.25$) and lanthanide-rich,
high-opacity ejecta (with $Y_e \lesssim 0.25$),
respectively~\citep{Arcavi2017, Chornock_etal2017, Tanvir2017}. Common
explanations for the neutron-rich ejecta in BNS mergers include tidal
disruption in asymmetric binaries and disk winds from prompt black hole
collapse or short-lived hypermassive neutron star (HMNS)
remnants~\citep{Siegel2017, Gill2019, Bernuzzi2024}. However, recent
studies suggest that a short-lived remnant alone cannot account for the
kilonova brightness and duration~\citep{Kawaguchi2023}. Accurate modeling
of the ejecta composition, light curve, and lifetime of the remnant
remains challenging given the difficulties of carrying out long-term
simulations~\citep{Kiuchi2023b, Ng2024b}, but, more importantly, due to
the uncertainties in the radiation transport and weak
interactions~\citep{Foucart2020c, Zappa2023, Cheong2024b, Foucart2024}.
Indeed, neutrino interactions significantly impact both ejecta
composition and the formation of short gamma-ray bursts~\citep{Ciolfi2020,
  Combi2023, Musolino2024b}, as well as the postmerger
dynamics~\citep{Radice2022, Foucart2024}.

To date, most numerical simulations of BNS mergers assume the neutron
stars to be made of neutrons ($n$), protons ($p$), electrons ($e^-$), and
positrons ($e^+$), \ie \npem. In this way, they neglect the presence of
muons ($\mu^-$), antimuons ($\mu^+$), and, obviously of muonic weak
interactions due to their high rest mass ($m_\mu = 105.7\, {\rm MeV}$).
While this is a choice dictated by simplicity, core-collapse supernova
simulations suggest that temperatures exceeding $\sim 30\, {\rm MeV}$ can
yield substantial muon production via electromagnetic and weak
interactions~\citep{Bollig2017}. Recent studies in simplified scenarios
have started to explore the role of muonic interactions in BNS mergers,
either using postprocessed muonic interactions~\citep{Loffredo2023} or
with \npemm but an approximate (\ie leakage-based) neutrino
treatment~\citep{Gieg2024}. In this Letter, we present the first
comprehensive analysis of BNS mergers in full general relativity with
\npemm and employing a moment-based treatment of neutrino transport with
five neutrino species.

\section{Numerical setup and neutrino microphysics}
We perform our simulations within the general-relativistic radiation
hydrodynamics (GRRHD) with adaptive mesh refinement provided by
the~\texttt{EinsteinToolkit} \citep{EinsteinToolkit_etal:2020_11}. The
GRRHD equations are solved with a fourth-order accurate finite differences
high-resolution shock-capturing scheme by the \texttt{FIL}
code~\citep{Etienne2015,Most2019b}. \texttt{FIL} provides its own
spacetime evolution code, \texttt{Antelope} based on the Z4 system
\citep{Bernuzzi:2009ex, Alic:2011a}, as well as a framework for handling
temperature and composition-dependent equations of state (EOSs)
\citep{Most:2018eaw, Most:2019onn}. High-quality initial data are obtained
with the \texttt{FUKA} solver~\citep{Papenfort2021b,
  Tootle2021}. \texttt{FIL-M1}~\citep{Musolino2023} provides an
energy-integrated, moment-based M1 scheme that has so far considered three
neutrino species (electron neutrinos $\nu_e$, electron antineutrinos
$\bar{\nu}_e$, and a collective species describing muon neutrinos $\nu_{\mu}$,
tau neutrinos $\nu_{\tau}$ and the corresponding antineutrinos $\nu_x$). As
anticipated, we have extended \texttt{FIL-M1} to evolve five neutrino species,
with the addition of $\nu_{\mu}$ and $\bar{\nu}_\mu$, while $\nu_{\tau}$ and
$\bar{\nu}_\tau$ are evolved as an effective species $\nu_x$ since
the $\beta$-processes involving them are negligible due to the high rest mass of
the tau particle ($1776.9\, {\rm MeV}$).

Extensions have been needed for the M1 module to self-consistently handle
the five neutrino species, and the GRRHD module has been enhanced with
the capability of a leptonic table accounting for pressure, internal
energy, and specific-entropy contributions to the nuclear EOS based on an
ideal Fermi gas of $\mu^-$ ($\mu^+$) and $e^-$ ($e^+$), where the
electron fraction $Y_e$ and muon fraction $Y_\mu$ are sampled within
$[0.01, 0.5]$ linearly and $[5\times10^{-4}, 0.2]$ logarithmically,
respectively. The quantities $Y_e$ and $Y_\mu$ are defined as $Y_l
  := Y^-_l - Y^+_l$, where $l \in {e, \mu}$, and $Y^-_l$ and $Y^+_l$
  represent the lepton and antilepton, respectively. In our simulations, 
  we ensure that $Y_l$ is positive as it is possible in BNS simulations
  that an excess of antileptons can lead to negative values of
  $Y_l$. This normally occurs only in highly localized regions and on
  very short timescales due to normal matter dominance and
  the charge-neutrality condition. Also, these excess antileptons can annihilate
  with leptons via rapid interactions, \eg $l^- + l^+ \rightarrow
  \gamma$, on timescales significantly shorter than those of weak
  interactions or our simulation time-stepping.

The leptonic table is then used in conjunction with the
conventional nuclear EOS table where the proton fraction ($Y_p = Y_e +
Y_\mu$) satisfies charge neutrality. Finally, the muon fraction is then
advected by \texttt{FIL-M1},
\begin{equation}\label{eq:mu_evol}
\partial_t (\sqrt{\gamma} \rho W Y_\mu) + \partial_i ( \sqrt{\gamma} \rho
u^i Y_\mu ) = - \mathcal{N}_{Y_\mu},
\end{equation}
where $u^\alpha$ is the fluid four-velocity, $W$ is the Lorentz factor, and
$\mathcal{N}_{Y_\mu} := m_{b}~(\mathcal{N}_{\nu_{\mu}} -
\mathcal{N}_{\bar{\nu}_{\mu}})$ is the collisional source term of the net
muon flavor neutrino number density, and $m_{b}$ is the rest mass of a
baryon. The conservative-to-primitive inversion and atmospheric treatment
routines were also updated to accommodate the additional degree of
freedom while ensuring lepton number conservation. The grid extent for
each simulation spans $[-1500 \, \rm{km}, 1500\,\rm{km}]^3$ with seven
refinement levels and a finest grid spacing of $\Delta x \approx 280\,
{\rm m}$~\citep{Tootle2022, Musolino2024b, Topolski2024c}.

\begin{table}[h!]
\centering
\begin{tabular}{ll}
\hline
 (a) $\nu_e+n \leftrightarrow p+e^{-}$ \;\;\;\;           & (e) $N+N \leftrightarrow N+N+\nu_{\mu/\tau}+\bar{\nu}_{\mu/\tau}$ \\
 (b) $\bar{\nu}_e+p \leftrightarrow n+e^{+}$  \;\;\;\;    & (f) $e^{-}+e^{+} \leftrightarrow \nu_{\mu/\tau}+\bar{\nu}_{\mu/\tau}$ \\
 (c) $\nu_\mu+n \leftrightarrow p+\mu^{-}$ \;\;\;\;       & (g) $\gamma_{\rm T} \leftrightarrow \nu_{\mu/\tau}+\bar{\nu}_{\mu/\tau}$ \\
 (d) $\bar{\nu}_\mu+p \leftrightarrow n+\mu^{+}$ \;\;\;\; & (h) $\nu_i+A \leftrightarrow \nu_i+A$  \\
 \;\;\;\; & (i) $\nu_i+N \leftrightarrow \nu_i+N $ \\
\hline
\end{tabular}
\caption{Note.~\cite{Ng2024a}. Here, $A$ refers to heavy nuclei, including light
    clusters such as $\alpha$-particles, and $^{2}H$, $N$ represents
    either a proton or a neutron, while $\nu_i$ is any neutrino species.}
\label{tab:weak_interactions}
\end{table}

For the neutrino microphysics, we have replaced the conventional
opacities used in \cite{Musolino2023} by the advanced ones provided by
the state-of-the-art neutrino microphysical library
\texttt{Weakhub}~\citep{Cheong2021, Cheong2023, Ng2024a}.  In addition to
standard neutrino interactions for \npem, we also include muonic
$\beta$-processes but exclude inelastic muonic interactions (\eg lepton
flavor exchange between $\mu^{-/+}$ and $e^{-/+}$) since our approach is
energy-integrated (gray approximation) and we lack neutrino energy
information. On the other hand, most of the corresponding
phase-space-integrated opacities of the inelastic muonic interactions are
significantly lower than the standard $\beta$-processes and elastic
scattering at high densities~(\eg see Figs.~(6.28) and (6.36) in
\citet{Lohs2015}).

{More specifically, we use the weak interactions listed in
  Table~\ref{tab:weak_interactions}}, which are computed by considering
an energy dependence using $18$ neutrino energy bins logarithmically
spaced in the range $[0.5,420]\,{\rm MeV}$ and tabulated after averaging
in energy following Eqs.~(A22)--(A23) in~\citet{Cheong2024b}. At runtime,
the opacities interpolated from the table and emissivities are
recomputed according to Kirchhoff's law with the rescaled degeneracy
parameters of the neutrinos (see Appendix~\ref{sec:strategy} for details
of the rescaled degeneracy parameters). For pair processes in both
  five- and three-species approaches that are relevant only for heavy-lepton 
neutrinos, \ie $\nu_{\mu,\tau}$ and $\bar{\nu}_{\mu,\tau}$ (see
    processes (e)-(g) in Table.~\ref{tab:weak_interactions}), we
calculate their isotropic emissivities on the fly
following~\citet{Ardevol2019} and apply Kirchhoff's law for the
corresponding absorption opacities. Note also that the pair processes
(e)-(g) are applied exclusively to heavy-lepton neutrinos and are
computed on the fly during the simulation using Eqs.~(B24)--(B32)
from~\citet{Ardevol2019}. Finally, we apply corrections to the
energy-averaged opacities by following Eqs.~(A24) and (A25)
in~\citet{Cheong2024b}. We note that when considering five species, 
  special treatments are needed for the opacity corrections and the
  limiting muonic interactions in some regimes (see
  Appendix~\ref{sec:strategy} for details).

We also note that in the three-species approach of \texttt{FIL-M1}, we
replace the emissivities and opacities with those calculated under weak
equilibrium while conserving lepton number and energy. This modification
was suggested by \citet{Radice2022} (see Eqs.~(90) and (91) therein) for
robustness and to accelerate the implicit solver by preventing
excessively small time-step sizes when the weak equilibration timescale
is too short to be resolved. This approach, however, cannot be employed
when considering five species because of the appearance of two distinct
equilibration timescales, \ie $npe\nu$ and $np\mu\nu$ equilibria; hence,
the Courant–Friedrichs–Lewy (CFL) factor for the five species 
is $0.2$, \ie half of that of used for three species.

Furthermore, due to the large rest mass of $\mu^-$, the use of an elastic
approximation for muonic $\beta$-processes~\citep{Oconnor2015,
  Musolino2023, Gieg2024}, can lead to significant discrepancies for the
muonic $\beta$-processes~\citep{Fischer2020b, Ng2024a}. Therefore, we
employ full kinematics calculations for $\beta$-processes to consider
momentum transfer~\citep{Fischer2020b, Guo2020, Ng2024a} with
self-consistent weak magnetism, nuclear-form factors, and medium
modifications in the nucleon propagator in dense
matter~\citep{Fischer2020b, Guo2020, Ng2024a}. Furthermore, reaction (g)
in Table.~\ref{tab:weak_interactions} accounts only for massive photons
(i.e., transverse electromagnetic modes that gain an effective mass
  by constantly interacting with $e^- e^+$ pairs in the
  plasma;~\citet{Ng2024a}), and the neutrino-nucleon elastic scattering
(i) incorporates nucleon recoil, weak magnetism~\citep{Horowitz2002},
and strangeness in the axial coupling ($g^s_A = -0.1$)~\citep{Hobbs2016}.

To establish the impact of muonic interactions on neutron star merger
remnants, we simulated four equal-mass and irrotational BNSs with a total
mass of $2.5 \, M_{\odot}$ and zero magnetic field to clearly isolate
pure muonic effects from the complex MHD dynamics. To assess the
robustness of the results, we employ two different realistic and
temperature-dependent EOSs, SFHo~\citep{Hempel2010} and
DD2~\citep{Typel:2009sy}, and for each EOS, we consider scenarios with
\npemm~(\ie five neutrino species and denoted as $5$-$\nu$ hereafter) or
with \npem~(\ie three neutrino species, or $3$-$\nu$). To compute the
initial conditions with the \texttt{FUKA} solver, we impose
zero-temperature, neutrino-less weak equilibrium~\citep{Gieg2024,
  Pajkos2024} and, to avoid double-counting of electron contributions, we
first subtract these from the table, then add different lepton
contributions according to whether we are considering $npe\mu$ or $npe$
matter. Finally, as useful diagnostic quantities, we introduce the
out-of-weak-equilibrium chemical potentials $\mu_{\Delta}^{npl} := \mu_n
- \mu_p - \mu_l$ (neutrinoless matter) and $\mu_{\Delta}^{npl\nu} :=
\mu_n - \mu_p - \mu_l - \mu_{\bar{\nu}_l}$ (neutrino-trapped matter),
where $\mu_i$ represents the chemical potential (including rest mass) for
species $n$, $p$, and leptons $l = e, \mu$, and where $\mu_{\nu_l}$ is
computed from the neutrino fraction $Y_{\nu_l}$~\citep{Espino2024b}.

\begin{figure}
  \includegraphics[width=0.5\textwidth]{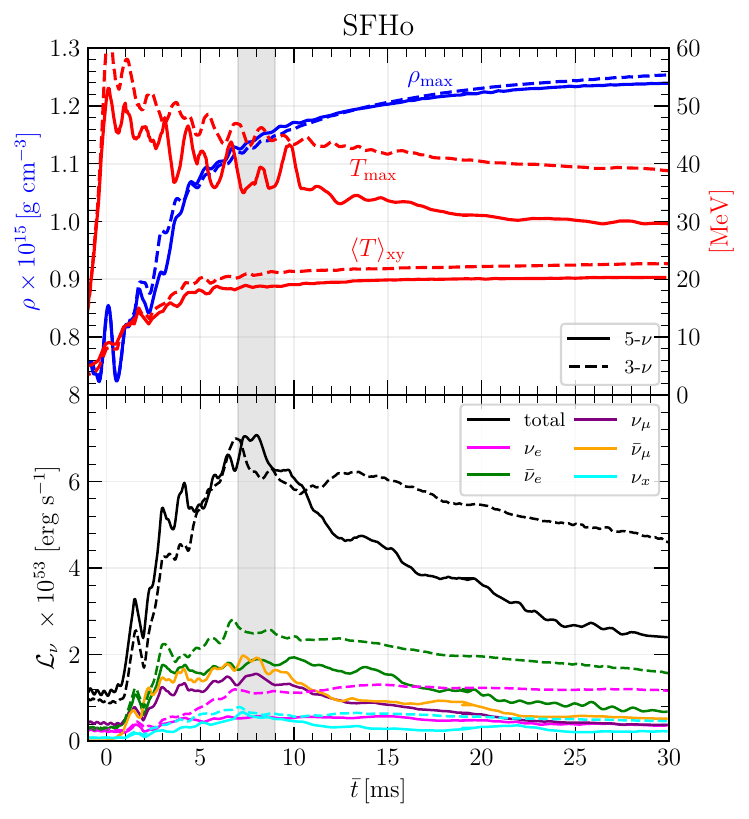}
  \caption{Top panel: evolution of the maximum rest-mass density
    $\rho_{\rm max}$, maximum temperature $T_{\rm max}$, and
    mass-averaged temperature $\langle T \rangle_{\rm xy}$ on the
    $(x,y)$-plane for the $5$-$\nu$ (solid lines) and $3$-$\nu$
    (dashed lines) scenarios. Bottom panel: neutrino luminosities (total
    and species-specific) $\mathcal{L}_{\nu}$ at $300\,{\rm km}$; the
    data is smoothed using a time-averaging over $0.2\, {\rm ms}$. In
    both panels, the shaded region denotes the time when
    $\mu^{npe\nu}_{\Delta} \approx \mu^{np\mu\nu}_{\Delta} \approx 0$
    (see gray shaded regions in Fig.~\ref{fig:mudelta_xy_avg} in
    Appendix~\ref{sec:dd2} for the SFHo and DD2 EOSs).}
  \label{fig:time_evolutions}
\end{figure}

\begin{figure*}
  \includegraphics[width=0.5\textwidth]{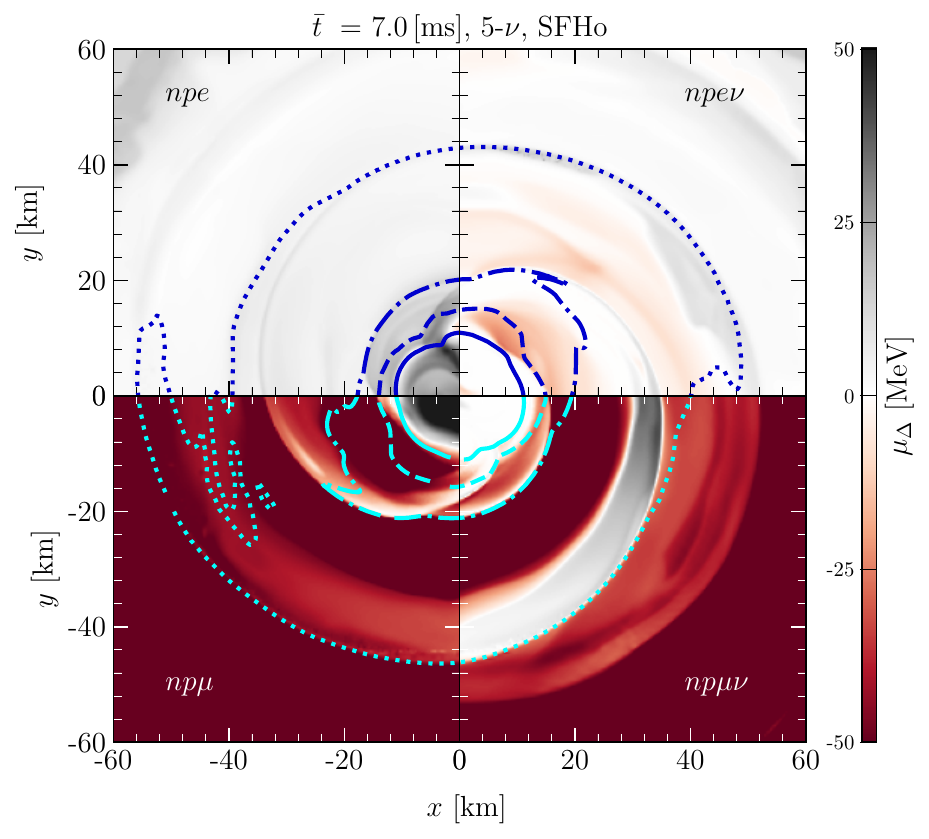}
  \includegraphics[width=0.49\textwidth]{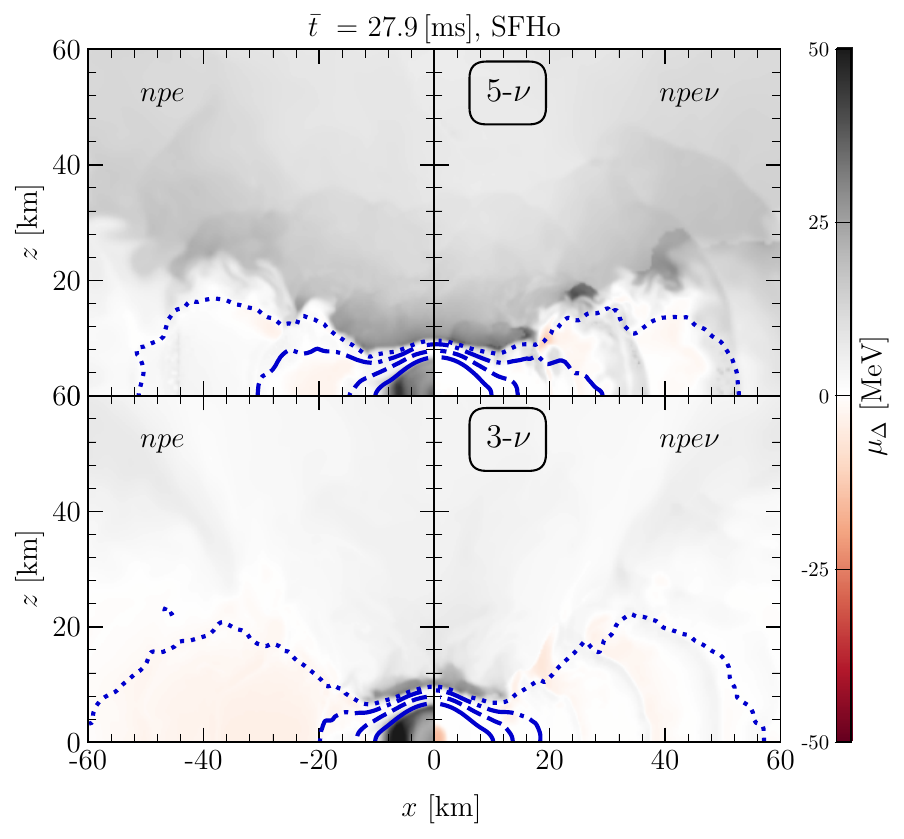}
  \caption{Left panel: equatorial distributions of the
    out-of-equilibrium chemical potentials $\mu^{npl}_{\Delta}$ (left
    portions) and $\mu^{npl\nu}_{\Delta}$ (right portions) for lepton
    flavors $l = e, \mu$ in the $5$-$\nu$ scenario at $\bar{t} = 7.0\,
    {\rm ms}$, \ie when $\langle \mu_{\Delta}^{npl\nu}\rangle_{\rm xy}
    \simeq 0$. The dotted, dashed-dotted, dashed, and solid lines show
    rest-mass density contours at $10^{11}, 10^{12}, 10^{13},
    \text{and}\, 10^{14}\, \rm g~cm^{-3}$, respectively. Right
      panel: the same as in the left panel, but at $\bar{t} = 27.9\,{\rm ms}$
    and on the $(x,z)$-plane. For compactness, we show distributions for
    $npe$ (left portions) and $npe\nu$ (right portions) equilibria but
    contrast the $5$-$\nu$ (upper part) with the $3$-$\nu$
    (bottom part) scenario. Note that $\mu_{\bar{\nu}_\mu} >
      \mu_{\bar{\nu}_e}$ in $5$-$\nu$ case (left panel) and that
      $\mu_{\bar{\nu}_e}$ is reduced in the $5$-$\nu$ case (right panel)
      because of the considerable differences in $\mu_{\bar{\nu}_e}$ in
      the high-density regions.} 
\label{fig:mudelta_beta}
\end{figure*}

\section{Results}

For compactness, the results presented hereafter will refer to the SFHo
EOS, but corresponding (and similar) results for the DD2 EOS can be found
in Appendix~\ref{sec:dd2}. The top part of Fig.~\ref{fig:time_evolutions}
illustrates the evolution of the maximum values of the rest-mass density
$\rho_{\rm max}$ (blue lines), temperature $T_{\rm max}$ (red lines), and
the temperature averages on the $(x,y)$-plane $\langle T \rangle_{\rm
  xy}$ (red lines) for the $5$-$\nu$ (solid lines) and the $3$-$\nu$
(dashed lines) scenarios. The data is shown as a function of the retarded
postmerger time $\bar{t} := t - t_{\rm mer}$, where $t_{\rm mer} \simeq
18.8\, {\rm ms}$ is the merger time~\citep{Rezzolla2016}, and differs by
$0.5\%$ in the two scenarios. Note that the $5$-$\nu$ scenario exhibits a
stronger cooling driven by several processes: larger neutrino emission
due to additional muonic processes, the generation of trapped neutrinos
via the conversion of matter internal energy into radiation internal
energy~\citep{Perego2019, Zappa2023}, and the creation of $\mu^-$-$\mu^+$
pairs from heating and the so-called ``muonization,'' \ie the process
where more $\bar{\nu}_{\mu}$ are emitted than $\nu_{\mu}$, thus building
up a net muon-lepton number~\citep{Bollig2017}. Muonization takes place
mostly in the shocked, dense, high-temperature, and sheared layers,
increasing the muonic fraction $Y_\mu$ to $\sim 0.04-0.05$ (see
Fig.~\ref{fig:2eoss} in Appendix~\ref{sec:dd2} for details). However, it
is considerably reduced as the matter moves outward, where it reaches
lower densities and lower temperatures and ``electronization'' occurs more
efficiently. An oscillatory behavior in $T_{\rm max}$ follows from
alternating muonization and demuonization in the hot spots of the two
stellar cores, which significantly alters the temperature by converting
or releasing electron degeneracy energy to form or absorb $\mu^-$ and
$\mu^+$ via muonic $\beta$-processes. In the $3$-$\nu$ case,
  (de)electronization does not alter the temperature significantly,
  leading to smaller oscillations in $T_{\rm max}$ due mostly to the
  oscillations of the remnant.

\begin{figure*}
  \includegraphics[width=1.\textwidth]{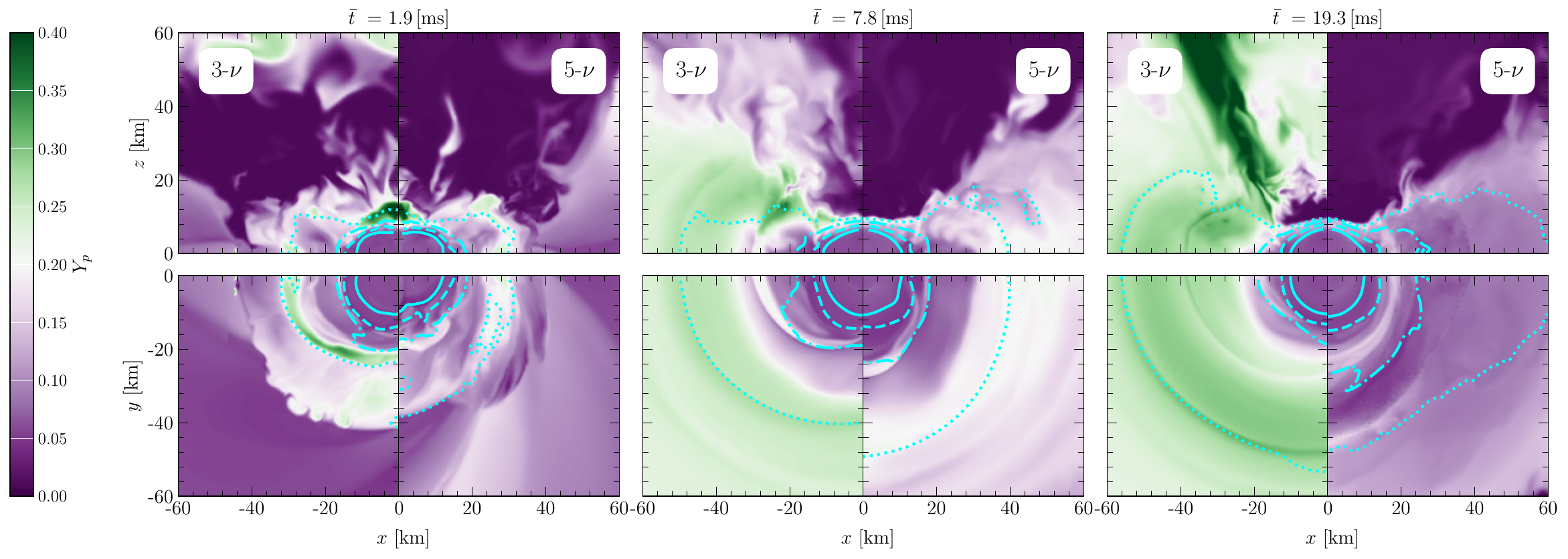}
  \caption{Distributions of the proton fraction $Y_p$ at $\bar{t} = 1.9\,
    {\rm ms}$ (left column), $7.8\, {\rm ms}$ ({middle} column), and
    $19.3\, {\rm ms}$ (right column). The top (bottom) row shows the
    polar (equatorial) distributions, and for each panel, the right
    (left) portions refer to the $5$-$\nu$ ($3$-$\nu$) scenarios,
    respectively. Note the differences that develop and the significantly
    lower protonization at late times when accounting for five neutrino
    species.}
  \label{fig:threemoments_sfho_yp}
\end{figure*}

The bottom part of Fig.~\ref{fig:time_evolutions} shows both the total
and the neutrino luminosities $\mathcal{L}_\nu$ across all species. In
the $5$-$\nu$ scenario, both neutron stars start with a central $Y_\mu
\approx 0.022$ and {$Y_e \approx 0.06$, while $Y_e \approx 0.07$ for
  the $3$-$\nu$} case~\citep{Loffredo2023}. The effects of the rapid
muonization can be deduced from the fact that during the first $8
\,\rm{ms}$ $\mathcal{L}_{\bar{\nu}_\mu}$ is $\sim 30\%$ larger than
$\mathcal{L}_{\nu_\mu}$ implying an excess of $\mu^-$ over $\mu^+$ in the
remnant. The emission of $\bar{\nu}_\mu$ ($\nu_\mu$) via $\mu^+$
($\mu^-$)-capture processes is able to extract a substantial amount of
energy from the remnant as $\mu^+$-capture processes tap energy that
would otherwise be used in $e^+$-capture. This obviously favors
$\mathcal{L}_{\bar{\nu}_{\mu}}$ over $\mathcal{L}_{\bar{\nu}_e}$ in
\npemm. Similarly, $\mu^-$-capture processes remove energy that would be
used for $e^-$-capture, thus resulting in a $\sim 12\%$ increase in the
total emitted neutrino energy within the first $8\, {\rm ms}$
postmerger. The larger neutrino luminosity ends at $\bar{t}
  \approx 8\, {\rm ms}$, after which it declines rapidly to be lower than
  that of the $3$-$\nu$ case. The start of this decline, which is marked
with a gray shaded area, corresponds to when the high-density matter
reaches a muonic trapped-neutrino weak equilibrium (or
$np\mu\nu$ equilibrium) and is characterized by having $\langle
\mu_{\Delta}^{np\mu\nu} \rangle_{\rm xy} \approx 0$ (see
Fig.~\ref{fig:mudelta_xy_avg} in Appendix~\ref{sec:dd2}), so that
$\mathcal{L}_{\nu_\mu}$ and $\mathcal{L}_{\bar{\nu}_\mu}$ are
significantly reduced. As to be expected, when this equilibrium starts to
be reached around $\bar{t} \sim 9\, {\rm ms}$, the oscillations in
$T_{\rm max}$ disappear. The stronger neutrino emission in the
early postmerger stages and the more efficient cooling -- via neutrino
emission, the generation of trapped neutrinos, $\mu^-$-$\mu^+$ pairs, and
via muonization -- observed in the $5$-$\nu$ scenario obviously leads to
a colder remnant and hence to a significant reduction in $\nu$ emission
across all species, resulting in a $\sim 50\%$ lower total
$\mathcal{L}_\nu$ at $\bar{t} = 30\, {\rm ms}$. It is worth
  pointing out that while the $3$-$\nu$ scenario tends to underestimate
  $\nu_\mu$ and $\bar{\nu}_\mu$ emissions by neglecting muonic
  $\beta$-processes, the $5$-$\nu$ scenario tends to underestimate
  $\nu_e$ and $\bar{\nu}_e$ emissions, although the latter is expected to
  be a smaller effect.

The process of $npl\nu$ equilibration can also be
appreciated also from the left panel of Fig.~\ref{fig:mudelta_beta},
which reports distributions of $\mu^{npl}_{\Delta}$ (left portions) and
$\mu^{npl\nu}_{\Delta}$ (right portions) at $\bar{t} = 7.0\, {\rm ms}$
for lepton flavors $l = e, \mu$ in the $5$-$\nu$ scenario. Note that
the system reaches $np\mu\nu$ and $npe\nu$ equilibria (\ie
$\mu^{npl\nu}_{\Delta} \simeq 0$) in regions with $\rho > 10^{14}\, {\rm
  g~cm^{-3}}$, where significant neutrino trapping (due to elastic
scattering and absorption processes) drives the system to recover the
broken $npe$ and $np\mu$ equilibria. These results, which are in
qualitative agreement with previous simplified
studies~\citep{Espino2024b}, indicate that the addition of new degrees of
freedom and muonic weak interactions still leads to weak equilibria a few
milliseconds after merger. The left panel of Fig.~\ref{fig:mudelta_beta}
also shows that $\mu_{\Delta}^{np\mu}$ exceeds $\mu_{\Delta}^{npe}$,
indicating that substantially more $\bar{\nu}_\mu$ are trapped compared
to $\bar{\nu}_e$. Additionally, the presence of nonzero $Y_\mu$ reduces
$\mu_e$~\citep{Bollig2017}, while $\mu^+$-capture partially replaces 
$e^+$-capture; hence, in the $5$-$\nu$ scenario, $\mu_{\bar{\nu}_e}$ is
decreased by $\sim 20-30\, {\rm MeV}$ (see right panel of
Fig.~\ref{fig:mudelta_beta}). Importantly, this phenomenology suggests
that the neutrino-trapping hierarchy normally encountered in simulations
with three neutrino species, \ie $\mu_{\bar{\nu}_e} > \mu_{\nu_x} >
\mu_{\nu_e}$~\citep{Zappa2023}, should be replaced by the new hierarchy
$\mu_{\bar{\nu}_\mu} > \mu_{\bar{\nu}_e} > \mu_{\nu_x} > \mu_{\nu_e} >
\mu_{\nu_\mu}$.

Following the merger, both electronic and muonic neutrino beta
  equilibria are achieved, \ie $\mu^{npe\nu}_{\Delta} \approx
  \mu^{np\mu\nu}_{\Delta} \approx 0$ and $\mu_{\bar{\nu}} \approx
  -\mu_{\nu}$.  As shown in the left panel of
  Fig.~\ref{fig:mudelta_beta}, $\mu^{np\mu}_\Delta$ exceeds
  $\mu^{npe}_\Delta$ in high-density regions in the $5$-$\nu$ scenario.
  Therefore, the system requires a higher value of $\mu_{\bar{\nu}_\mu}$
  than $\mu_{\bar{\nu}_e}$ to satisfy the conditions above. In addition,
  this also naturally implies that $\mu_\mu < \mu_e$, \ie that muons are
  less degenerate than electrons. This smaller degree of degeneracy of
  muons in dense matter, combined with the requirement for neutrino beta
  equilibrium, is key to explaining the new hierarchy in the neutrino
  fraction $Y_\nu$, that is, $\mu_{\bar{\nu}_\mu} > \mu_{\bar{\nu}_e} >
  \mu_{\nu_x} > \mu_{\nu_e} > \mu_{\nu_\mu}$. Indicating the neutrino
  fractions as $Y_{\nu}$ and recalling that $Y_{\nu} \sim {T^3
    F_2(\mu_{\nu})}/{\rho} $, with $F_2(\mu_{\nu})$ the second-order
  Fermi integral, the hierarchy can also be expressed as
  $Y_{\bar{\nu}_\mu} > Y_{\bar{\nu}_e} > Y_{\nu_x} > Y_{\nu_e} >
  Y_{\nu_\mu}$.

\begin{figure*}
\includegraphics[width=0.99\textwidth]{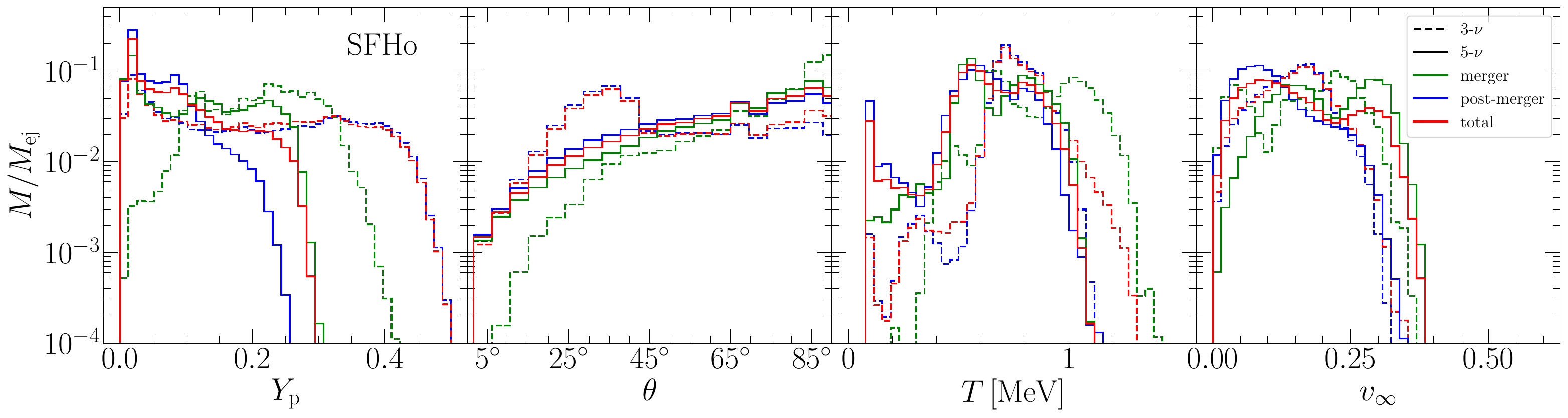}
\caption{Distributions of the normalized ejected mass for the $5$-$\nu$
  (solid lines) and $3$-$\nu$ (dashed lines) scenarios as functions
  (from left to right) of the proton fraction $Y_p$, the polar angle
  $\theta$, the temperature $T$, and the terminal velocity
  $v_{\infty}$. The data are extracted on a two-sphere detector at $300\,\rm{km}$.
  Shown with red lines are the total distributions, while green and blue
  lines refer respectively to $\bar{t} \leq 4.0\,\rm{ms}$ and $\bar{t}
  \geq 4.0\,\rm{ms}$ (different times reflect the different speeds of the
  ejecta to reach the detector). In all cases, $Y_p \approx Y_e$, since
  $Y_{\mu} \lesssim 5\times 10^{-3}$.}
\label{fig:histogram_sfho}
\end{figure*}

Special attention has been paid to assess how the inclusion of five
neutrino species changes the composition of the remnant matter and the
properties of the unbound ejecta, that is, matter with $-h u_t > 1$,
where $h$ is the specific enthalpy and $u_t$ is the covariant time
component of the fluid four-velocity~\citep{Bovard2016}.
Figure~\ref{fig:threemoments_sfho_yp} is meant to illustrate the changes
in the composition at three different times ($\bar{t}\simeq 2, 8$, and
$20\,{\rm ms}$), either in the polar plane (top row) or in the equatorial
one (bottom row) and for both the $3$-$\nu$ (left part of each column)
and for the $5$-$\nu$ (right part) scenarios. Soon after merger (left
column), the protonization is weaker in the $5$-$\nu$ scenario, as energy
is redirected toward the creation of $\mu^-$ and $\mu^+$ and muonic
processes (muonization indirectly converts $Y_e$ to $Y_\mu$, thus
reducing $\mu_e$~\citep{Bollig2017}). The net changes in $Y_p := Y_e +
Y_\mu$ are due to differences in the rates of antilepton/lepton capture
and, although the differences in $\mathcal{L}_{\bar{\nu}_\mu}$ and
$\mathcal{L}_{\nu_\mu}$ are small, these rates deprive the system of a
significant amount of energy that would be spent in emitting
${\bar{\nu}_e}$ and ${\nu}_e$, thereby reducing the gap between
$\mathcal{L}_{\bar{\nu}_e}$ and $\mathcal{L}_{\nu_e}$ and inducing a
weaker electronization. Furthermore, demuonization occurs quickly in the
ejecta, reducing $Y_\mu$ to $\lesssim 10^{-3}$ and resulting in dynamical
ejecta that are less proton-rich, with $Y_p \approx Y_e$; this will have
important nucleosynthetic consequences (see below).

These trends in neutrino luminosities persist throughout the simulation
and become more pronounced at later times, as $\nu_e$ and $\bar{\nu}_e$
emissions are significantly suppressed, not only due to the colder
remnant but also because of the additional energy cost associated with
muonic processes. This is supported by the observation that, despite the
lower remnant temperature, for $\bar{t} \lesssim 12\, {\rm ms}$ the
combined heavy-lepton neutrino luminosity is larger in the $5$-$\nu$
scenario, \ie $\mathcal{L}_{\nu_\mu} + \mathcal{L}_{\bar{\nu}_\mu} + 2
\mathcal{L}_{\nu_x}\vert_{5{\text{-}}\nu} > 4\mathcal{L}_{\nu_x}
\vert_{3{\text{-}}\nu}$, where the luminosities on either side of the
inequality refer to the five or three neutrino species, respectively.
Moreover, the disk (\ie the portion of the remnant with $\rho \lesssim
10^{12}\,{\rm g~cm}^{-3}$) is cooler in the $5$-$\nu$ case, which leads
to a lower absorption opacity for $\nu_e$ and a weaker $\nu_e$
reabsorption. Additionally, the different hierarchy of trapped neutrinos
reduces $\mu_{\bar{\nu}_e}$ in a shell-like region of high
temperature~\citep{Hanauske2016}, resulting in reduced
$\eta_{\bar{\nu}_{e}}$. This weakens the $e^+$-capture to produce $p$ and
$\bar{\nu}_e$ in the polar region. These factors significantly weaken
protonization in the disk and polar regions as shown in the middle and
right columns of Fig.~\ref{fig:threemoments_sfho_yp} (right portions). By
contrast, the $3$-$\nu$ scenario features stronger electronization in the
hotter disk that is driven by a stronger $\nu_e$ reabsorption and 
$e^+$-capture (left portions). Finally, since $\mu_{\Delta}^{np\mu} \lesssim
-50\, {\rm MeV}$ in regions with $\rho \lesssim 10^{14}\, {\rm
  g~cm^{-3}}$ (see Fig.~\ref{fig:mudelta_beta}), $\mu^-$ are unlikely to
survive in the comparatively low-density, low-temperature regions, thus
leading to stronger (weaker) $\mu^-$ ($\mu^+$)-capture, \ie to a net
demuonization ($Y_\mu/Y_p \sim 10^{-4}-10^{-2}$) in the disk and at
high latitudes. Overall, our simulations reveal that a competition
between muonic $\beta$-processes and electronic $\beta$-processes further
weakens the protonization and leads to a more neutron-rich disk and
high-latitude wind in the $5$-$\nu$ scenario. However, because
$\mu_{\Delta}^{npe} > 0$ in the disk and in the high-latitude regions (see
right panel of Fig.~\ref{fig:mudelta_beta}), gradual electronization is
still occurring, albeit at a significantly slower rate than in a
$3$-$\nu$ scenario.

As a concluding aspect of our analysis, we study the composition of the
unbound ejecta. More specifically, we follow \citet{Radice2016,
  Bovard2016} and \citet{Radice2018a}, binning the unbound fluid elements with a
$40\times40$ grid in specific entropy $s$ and $Y_p$, where the weight is
given by the accumulated mass fraction. The median unbound fluid element
is sampled from each nonempty bin, leading to an average number of
$\approx 1000$ ``tracers''. We determine the dynamical timescale of the
expansion of ejecta, $\tau$, by equating the homologous evolution
$\rho(t) = \rho_{\rm ext} (v_{\rm ext}t/r_{\rm ext})^{-3}$, where
$\rho_{\rm ext}$ and $v_{\rm ext}$ are the density and velocity of the
fluid element when it crosses the radius of the 2-sphere detector $r_{\rm
  ext}$ at time $t$, with that obtained considering an expanding material
that undergoes $r$-process nucleosynthesis, $\rho(t) = \rho_0
(3\tau/\exp(1)\,t)^3$, where $\rho_0=\rho_0(s, Y_e, 6\,{\rm
  GK})$~\citep{Lippuner2015}. Over a timescale of $\bar{t} = 28\, {\rm
  ms}$ we measure a total amount $\approx 1.63 \times 10^{-3}\,
M_{\odot}$ ($\approx 3.64 \times 10^{-3} \, M_{\odot}$) for the $5$-$\nu$
($3$-$\nu$) scenario, respectively. The factor-two difference in the
ejected matter in the $5$-$\nu$ simulations is due to the combination of
a larger neutrino luminosity right after merger -- that deprives the
system of kinetic energy -- and of a consequent cooler remnant -- that
launches weaker neutrino-driven winds. Furthermore, while we measure a
negligible fraction of muons ($Y_\mu \lesssim 5 \times 10^{-3}$), ejecta
with low (high) $Y_p$ is significantly increased (suppressed).

Figure~\ref{fig:histogram_sfho} shows the ejecta properties as
distributions of the normalized ejected mass for the $5$-$\nu$ (solid
lines) and $3$-$\nu$ (dashed lines) scenarios, as functions of (from left
to right) the proton fraction $Y_p$, the polar angle $\theta$, the
temperature $T$, and the terminal velocity $v_{\infty}$. Shown with red
lines are the total distributions, while green and blue lines refer
respectively to $\bar{t} \leq 4.0\,\rm{ms}$ and $\bar{t} \geq
4.0\,\rm{ms}$ to distinguish the properties around the merger from those
in the postmerger. Overall, we find that differences emerge between the
$5$-$\nu$ and $3$-$\nu$ scenarios, with the muonic merger yielding around
twice the fraction of neutron-rich ejecta in the range $Y_p \sim
0.05$--$0.1$, suppressing proton-rich ejecta {($Y_p \gtrsim 0.3$)},
and producing ejecta that have lower $T$, and are less concentrated at
high latitude ($\theta < 45^{\circ}$).

The $r$-process yields are then computed using the nuclear-reaction
network code \texttt{SkyNet}~\citep{Lippuner2017}, using a workflow
similar to that in~\citet{Papenfort2018} and the rates in
\citet{Cyburt2010, Panov2010, Roberts2011}. The assumed nuclear
statistical equilibrium is computed by \texttt{SkyNet} on the basis of
the values of $T$, $s$, and $Y_p$ of each tracer, where the rest-mass
density evolution consists of an exponential decrease transitioning to an
homologous expansion for $t \geq 3\, \tau$, {where $\tau$ is the
  timescale of the expansion of ejecta discussed
  above~\citep{Lippuner2015}. Hence, for each simulation, we provide
  \texttt{SkyNet} with information on $T$, $s$, $Y_p$, and $\tau$.}

\begin{figure}
  \hspace*{-0.95em}
  \includegraphics[width=1.0\columnwidth]{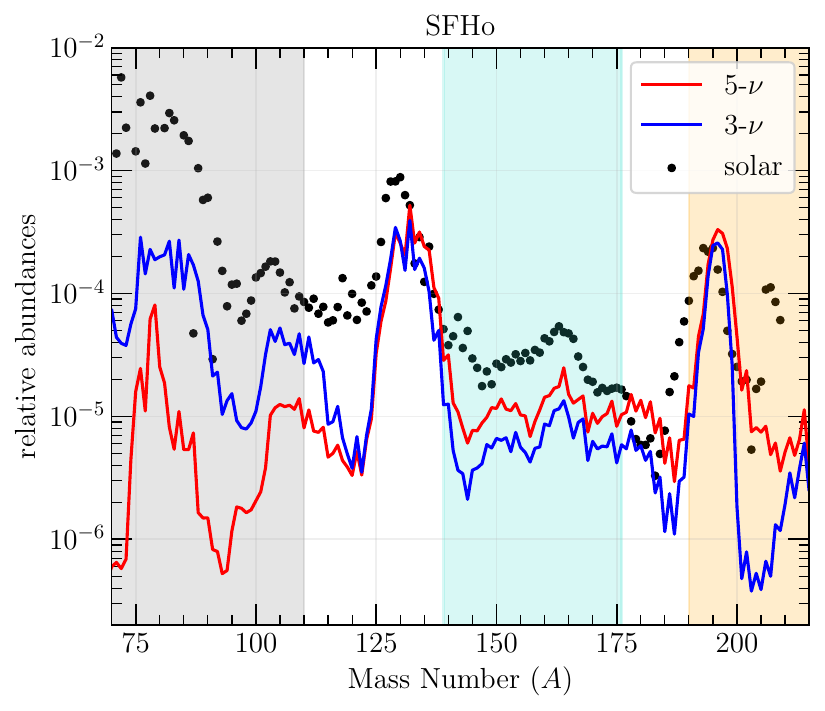}
  \caption{Comparison between the solar relative abundances (black filled
    circles) as a function of the mass number $A$ and the nucleosynthetic
    yields from the $5$-$\nu$ (red line) and the $3$-$\nu$ (blue line)
    scenarios; both abundances are rescaled to match at $A = 132$. Note
    the better match with data for lanthanides $139 < A < 176$
    (blue shaded region) and for very heavy elements with $190 < A < 215$
    (orange shaded); very light elements with $70 < A < 110$ (gray
    shaded) are significantly suppressed.}
\label{fig:abundance}
\end{figure}

Figure~\ref{fig:abundance} reports the $r$-process yields for the
$5$-$\nu$ scenario (red line) and for the $3$-$\nu$ scenario (blue line),
comparing them to the abundances measured in the solar system (filled
black circles). When comparing the two scenarios, it is apparent that the
inclusion of five neutrino species leads to larger abundances of
lanthanides, \ie with mass number $139 < A < 176$ (blue shaded region),
and to almost an order of magnitude larger yields for elements with $190
< A < 215$ (orange shaded region)\footnote{We also find an $18\%$ better
match in the actinides range ($232 < A < 238$) not reported in
Fig.~\ref{fig:abundance} because of the large uncertainty in the solar
data.}. At the same time, elements with $70 < A < 110$ are significantly
suppressed (gray shaded shaded).  Hence, our results for the SFHo EOS
clearly indicate that inclusion of muonic interactions provides a $\sim
100\%$ increase in the yields of lanthanides.

It has been noted that the red component of a kilonova emission is
  considered too massive to be attributed solely to dynamical ejecta, and
  that the blue component is too fast to be coming from the ejecta in the
  polar region of the long-lived remnant~\citep[see, \eg][]{Kawaguchi2018b,
    Anand2023, Kawaguchi2023}. Different models have been proposed to
  address these tensions. In particular, the blue component has been
  interpreted as originating preferentially from ejecta produced by a
  short-lived HMNS, while the red component is thought to be
  predominantly equatorial due to neutrino shielding by the accretion
  disk~\citep{Metzger2017, Kasen2017, Curtis2023}. The results of our
  $5$-$\nu$ simulations provide an alternative and possibly natural
  explanation and hint at a scenario in which a long-lived HMNS with
  muonic interactions suppresses the overproduction of proton-rich
  matter, hence removing the need for an excessively rapid blue component
  from the polar regions.

While promising, our findings also come with a caveat. Specifically,
simulations carried out with the stiffer DD2 EOS, which is only
marginally consistent with astronomical constraints~\citep{Most2018},
indicate that the differences in the lanthanide yields are much smaller
(see Fig.~\ref{fig:abundance_dd2} in Appendix~\ref{sec:dd2}). This
different behavior is rooted in many factors: lower temperatures and
densities producing weaker shocks for the DD2 EOS~\citep{Espino2024a},
significantly less ejected matter and weaker muonization, and, more
importantly, a smaller reduction of $Y_p$ in postmerger, at least over
the timescales explored. We conjecture that muonic interactions may
require longer timescales to boost lanthanide production in stiff EOSs
and plan to explore such scenarios in future works.

\section{Conclusions}

We have presented the first series of GRRHD simulations of BNS mergers
employing a moment-based, energy-integrated approach to describe the
radiative transport via five neutrino species with $\mu^-$, $\mu^+$ and
muonic $\beta$-processes based on full kinematic calculations and medium
modifications in the nucleon propagator in dense matter.

Our results reveal that while an $np\mu\nu$ equilibrium, similar to an
$npe\nu$ equilibrium, is established within a few milliseconds
postmerger, the unique properties of muons and out-of-weak-equilibrium
potentials are such that a significant demuonization occurs in
low-density, low-temperature regions after the merger. The creation of
$\mu^-$-$\mu^+$ pairs and the muonization in the heated regions right
after the merger, together with a more intense initial neutrino emission,
lead to a cooler remnant and disk. Muonic-interaction channels require a
significant amount of energy, which would be otherwise tapped by
electronic $\beta$-processes to drive protonization of the ejecta and
disk, leading to an overall reduction of $\nu_e$-reabsorption and
$\bar{\nu}_e$ emission. Together with the changed hierarchy of trapped
neutrinos and competition between muonic and electronic processes, these
factors result in a weaker protonization and a more neutron-rich ejection
of matter over dynamical timescales. In turn, this boosts lanthanide
production and suppresses the formation of lighter elements via
$r$-process nucleosynthesis, {thus providing a natural possible
  explanation for} the overproduction of proton-rich matter in the
standard three-species scenario.

While being the first calculations of this type, our simulations also
come with some limitations that will need to be addressed in future
works. First, they neglect the presence of a dynamically important
magnetic field that could influence the dynamics of the remnant and of
the ejecta~\citep{Ciolfi2020_a, Moesta2020, Chabanov2022, Most2023,
  Combi2023, Kiuchi2023, Bamber2024, Aguilera-Miret2024,
  Jiang2025}. As a result, the properties of the outflows discussed
  here may be particularly sensitive to neutrino interactions and could
  change when magnetically driven winds are properly taken into account.
  Second, the use of equal-mass binaries implies that the dynamical ejecta
  and the disk mass computed here represent lower limits of what should
  be expected in more realistic unequal-mass conditions. Third, without
a spectral transport~\citep{Foucart2020c}, leptonic weak interactions
converting $\mu^{-/+}$ to $e^{-/+}$~\citep{Lohs2015, Fischer2020b,
  Guo2020}, the realistic spectral blocking for pair processes and
kernel-based processes~\citep{Bollig2017, Cheong2024c} are unaccounted,
and the use of a soft EOS maximizes (de)muonization, our approach may
slightly overproduce heavy elements. Finally, the evidence of positive
values for the out-of-weak-equilibrium chemical potential
$\mu_{\Delta}^{npe}$ indicates that electronization could take place on
timescales much longer than those explored here.

\bigskip
\section{Acknowledgments}

It is a pleasure to thank L. Boccioli, R. Bollig, F. Francois,
T. Fischer, T. Janka, E. Most, and D. Radice for useful
discussions. Support comes from the State of Hesse within the Research
Cluster ELEMENTS (Project ID 500/10.006) and through the European
Research Council Advanced Grant ``JETSET: Launching, propagation and
emission of relativistic jets from binary mergers and across mass
scales'' (grant No. 884631). LR acknowledges the Walter Greiner
Gesellschaft zur F\"orderung der physikalischen Grundlagenforschung
e.V. through the Carl W. Fueck Laureatus Chair. ST acknowledges support
from NASA award ATP-80NSSC22K1898. The calculations were performed in
part on the local ITP Supercomputing Clusters Iboga and Calea and in part
on HPE Apollo HAWK at the High Performance Computing Center Stuttgart
(HLRS) under the grants BNSMIC and BBHDISKS.

\begin{figure*}
\includegraphics[width=0.95\textwidth]{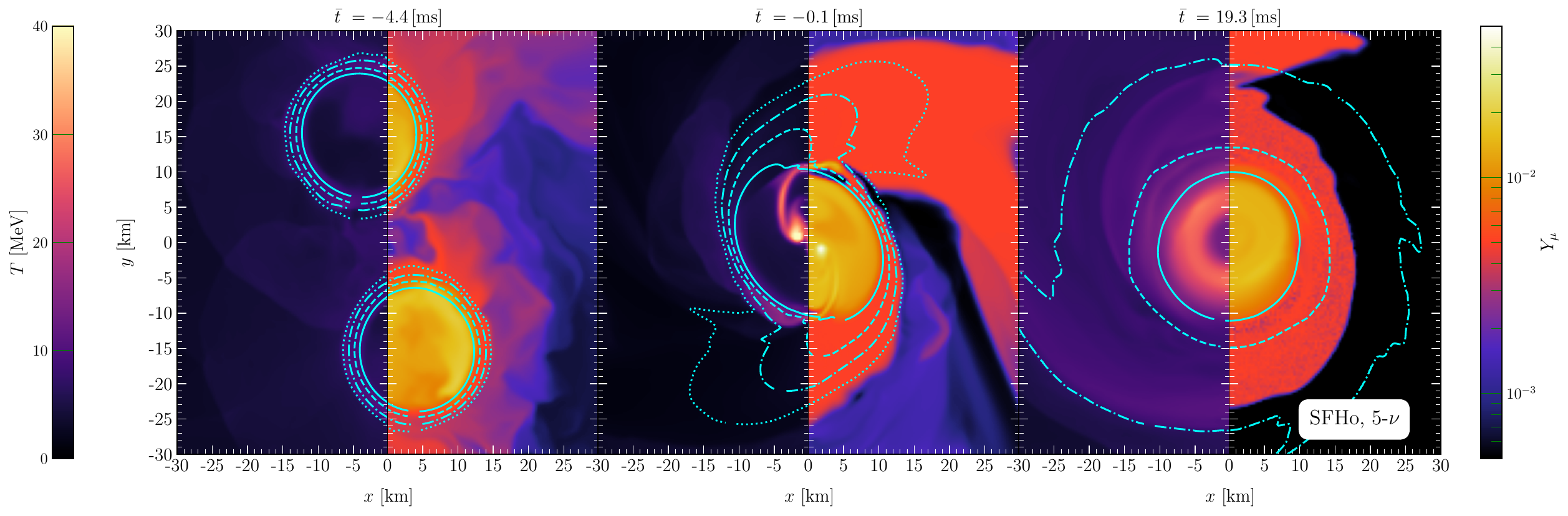}
\includegraphics[width=0.95\textwidth]{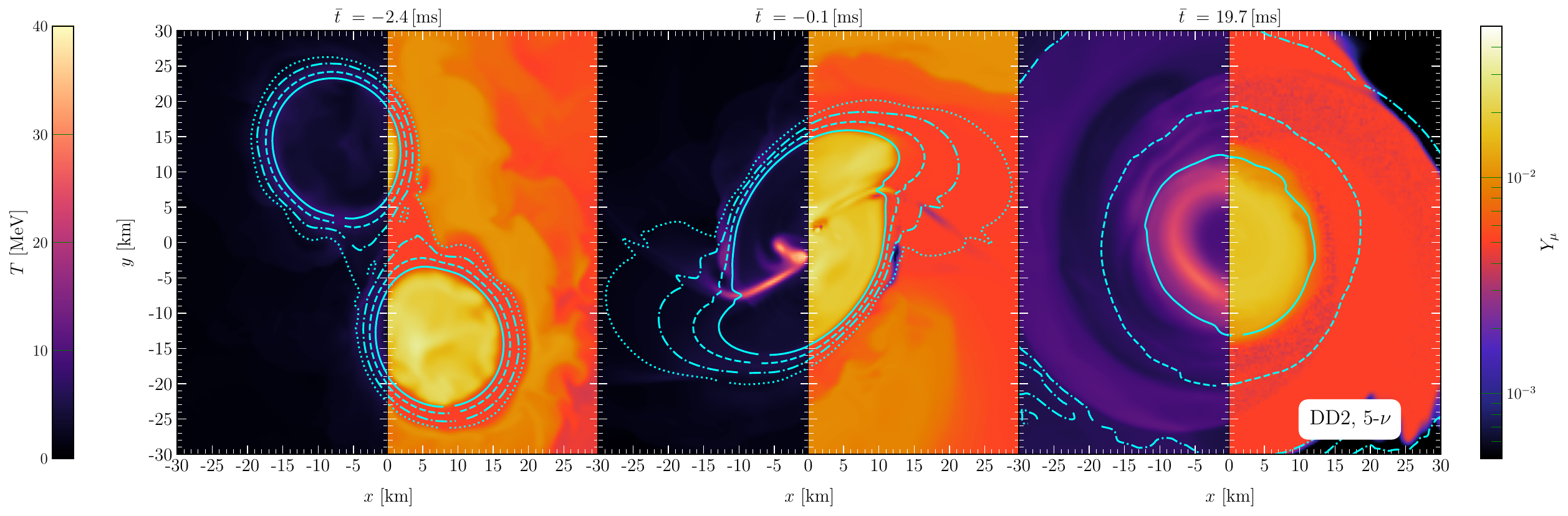}
      \caption{Columns, left to right: qquatorial snapshots of
      the $5$-$\nu$ scenarios for the SFHo (upper) and DD2
      (bottom) at the inspiral, merger, and postmerger phases,
      where left (right) of each panel is associated
      with $T$ ($Y_\mu$). The dotted, dashed-dotted, dashed, and solid
      lines show rest-mass density contours at $10^{11}, 10^{12},
      10^{13}, \text{and}\, 10^{14}\, \rm g~cm^{-3}$, respectively,
      highlighting the dependence of muonization on high density and
      temperature.}
        \label{fig:2eoss}
\end{figure*}

\newpage
\section*{Appendix}

\section{Impact of the EOS}
\label{sec:dd2}

Here we provide supplementary information about the results of the
simulations that either complement that provided in the main text for the
SFHo EOS or contrast the results when considering binaries evolved with
the DD2 EOS~\citep{Hempel2010}. We start by comparing and contrasting the
EOS influence on (de)muonization using Fig.~\ref{fig:2eoss}, which
reports equatorial distributions of the temperature $T$ (left portions)
and muon fraction $Y_{\mu}$ (right portions) during the inspiral (left
column), merger (middle column), and postmerger (right column) phases of
a $5$-$\nu$ scenario with the SFHo (top row) and the DD2 (bottom row)
EOS. During the inspiral, we measure values $Y_\mu \approx 0.022$ ($\approx 0.032$) at the stellar
centers for the SFHo (DD2) EOSs, which are essentially determined by the
neutrinoless $\beta$-equilibrium and negligible muonic interactions. Note
that the higher central muon fraction ($Y_\mu \approx 0.032$) for the DD2
EOS in neutrinoless $\beta$-equilibrium is essentially determined by the
symmetry energy of the EOS~\citep{Loffredo2023}. However, rapid
muonization at merger drives $Y_\mu$ to values $\simeq 0.036-0.048$
($\simeq 0.035$--$0.040$) for the SFHo (DD2) EOS. Subsequently, a gradual
demuonization takes place during postmerger, which is attributed to the
slowly cooling remnant and disk over the dynamical timescale. In this
phase, muons are present mostly in the hot shell clearly visible in the
temperature distribution.

To analyze the global behavior of both demuonization and muonization, we
compute the conserved muon number
\begin{equation} 
  N_\mu
  := \frac{1}{m_{b}}\int_{\mathcal{V}} \sqrt{\gamma} \rho W Y_\mu d^3x, 
\end{equation} 
where $\mathcal{V}$ is the simulation domain.

\begin{figure}
\hspace{-0.5cm}
  \includegraphics[width=0.5\textwidth]{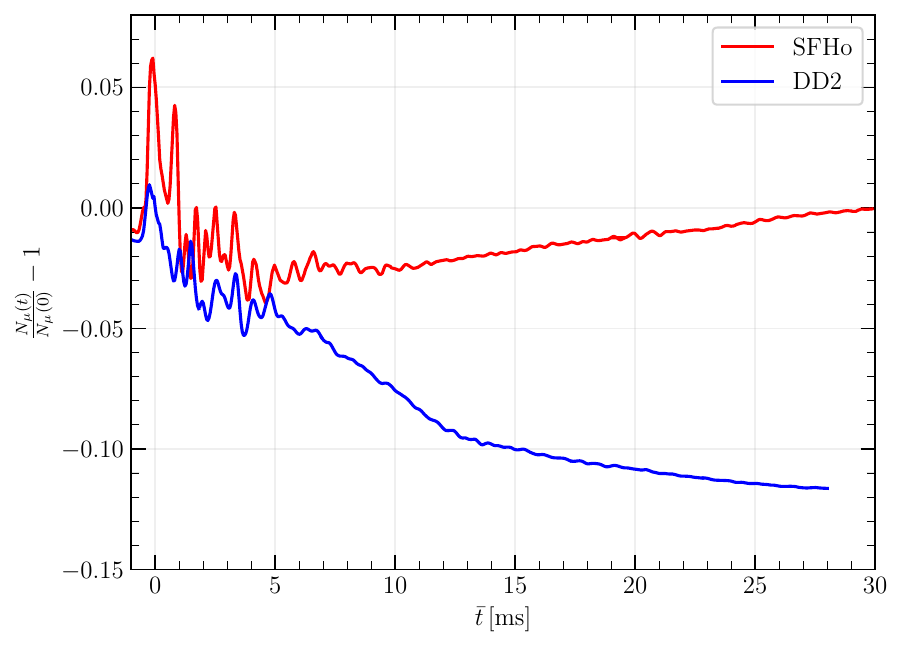}
  \caption{Evolution of the conserved muon number $N_\mu(t)$
      normalized to the initial value $N_\mu(0)$ for the two EOSs
      considered.}
  \label{fig:Nmu}
\end{figure}

Figure~\ref{fig:Nmu} shows the evolution of the conserved muon
  number $N_\mu(t)$ normalized to the initial value $N_\mu(0)$ for the
  two EOSs considered. During the inspiral phase, artificial heating from
  numerical perturbations and shock heating at the interface between the
  stellar surfaces and atmosphere lead to slight demuonization at the
  stellar surfaces (see also~\cite{Gieg2024}). Consequently, $N_\mu(t)$
  is slightly smaller than $N_\mu(0)$ for $\bar{t} <
  -1\,\mathrm{ms}$. Just before the merger, the stellar surfaces and
  tidally disrupted stellar material come into contact, triggering rapid
  muonization due to a sudden increase in density and
  temperature. Subsequently, as the remnant oscillates violently, it
  experiences rapid demuonization in relatively lower-density regions
  and oscillations in the conserved muon number as it expands,
  oscillates, cools, and ejects matter. At the same time, muonization
can still take place in the remnant core and surrounding regions through
the reabsorption of $\nu_\mu$. Note that although the amount of
  demuonization after the merger at $\bar{t}\simeq 0$ appears to be
  stronger in the case of the SFHo EOS, it is actually comparable to that
  in the DD2 EOS when compared to the maximum value of $N_\mu(t)$. After
  $\bar{t} \approx 5\,\rm{ms}$, the conserved muon number for the SFHo EOS
  stabilizes and begins to increase gradually, whereas for DD2 EOS, it
  continues to decrease, albeit at a lower rate.

The evolution of the conserved muon number can also be inferred
  from the neutrino average energy. In the energy-averaged moment scheme,
  in fact, the average neutrino energy is computed as
  $\langle~\epsilon_\nu~\rangle~=~W (\bar{E}_\nu - \bar{F}^i_\nu
  v_i)/N_\nu$ where $\bar{E}_\nu$ and $\bar{F}^i_\nu$ represent the
  averaged zeroth and first moments of radiation, respectively, and
  $N_\nu$ is the neutrino number density. For the SFHo EOS,
  $\mathcal{L}_{\bar{\nu}_\mu} > \mathcal{L}_{\nu_\mu}$, and the
  conserved muon number remains constant or slightly
  increases. Consequently, $\langle \epsilon_{\bar{\nu}_\mu} \rangle$ is
  significantly larger than $\langle \epsilon_{\nu_\mu} \rangle$. In
  contrast, for the DD2 EOS, $\mathcal{L}_{\bar{\nu}_\mu} <
  \mathcal{L}_{\nu_\mu}$, and the conserved muon number decreases
  monotonically, resulting in a smaller difference between $\langle
  \epsilon_{\bar{\nu}_\mu} \rangle$ and $\langle \epsilon_{\nu_\mu}
  \rangle$.

Clearly, the evolutions shown in Fig.~\ref{fig:Nmu} combine both
muonizations and demuonizations that depend sensitively on the region of
the remnant, on the EOS (degeneracies of electrons and muons), and
weak interactions rates, all of which are still poorly
constrained. These changes in the conserved muon number are also reflected
also in the difference between the $\mathcal{L}_{\nu_\mu}$ and
$\mathcal{L}_{\bar{\nu}_\mu}$ neutrino luminosities shown in
Figs.~\ref{fig:time_evolutions} and \ref{fig:time_dd2}.

\begin{figure}
\centering
  \includegraphics[width=0.45\textwidth]{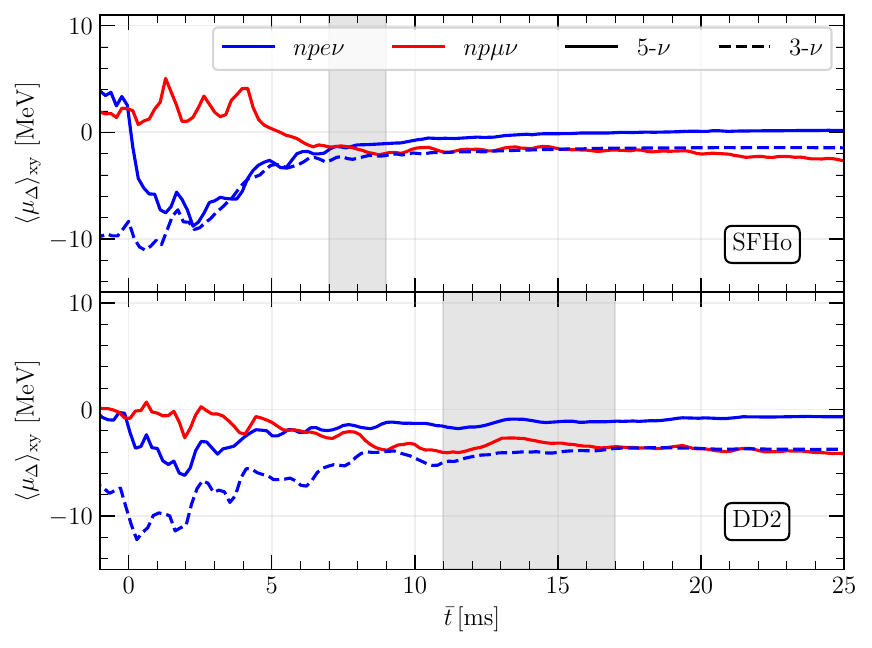}
\caption{Evolutions of the mass-averaged out-of-weak-equilibrium
  potentials $\langle\mu_{\Delta}\rangle_{\mathrm{xy}}$ for the $3$-$\nu$
  (dashed lines) and $5$-$\nu$ scenarios (solid lines) when employing the
  SFHo EOS (top panel) or the DD2 EOS (bottom panel). Both $npe\nu$ and
  $np\mu\nu$ equilibria with asymptotic values $\mu^{npe\nu}_{\Delta}
  \approx \mu^{np\mu\nu}_{\Delta} \approx 0$ in the gray shaded regions
  that are reported also in Fig.~\ref{fig:time_evolutions} and
  Fig.~\ref{fig:time_dd2}.  Note that $\mu_\nu$ for calculating
  $\mu_{\Delta}^{npl\nu}$ is most accurate in regions with $\rho >
  10^{12}~\rm{g~cm^{-3}}$~\citep{Espino2024b} where the neutrinos are
  equilibrated thermally with the matter, 
  and the mass-averaged values of $\langle\mu_{\Delta}^{npl\nu}\rangle_{\mathrm{xy}}$ 
  are determined mainly in the high-density regions.}
 \label{fig:mudelta_xy_avg}
\end{figure}

\begin{figure}
\includegraphics[width=0.45\textwidth]{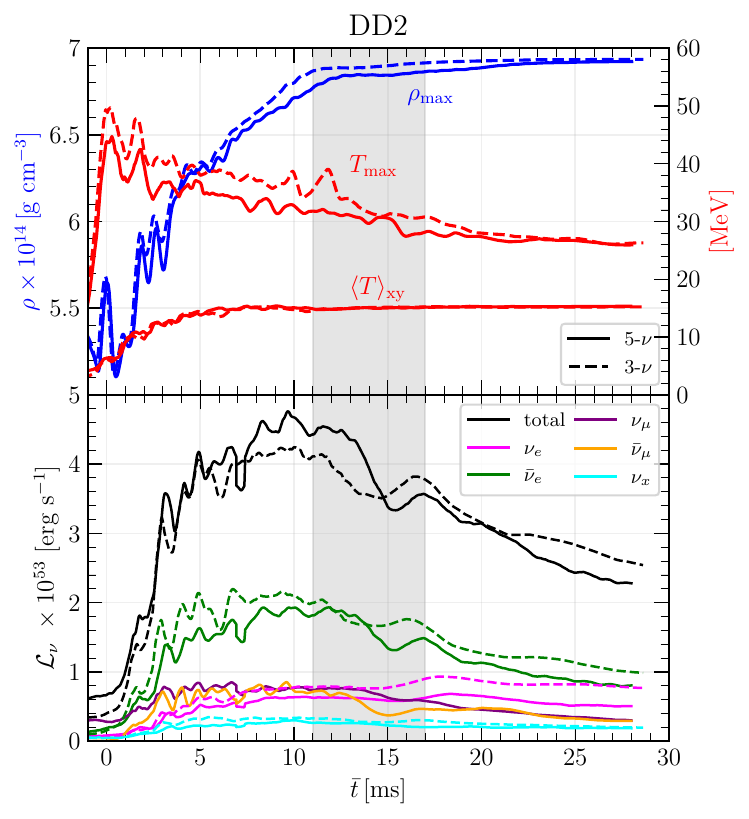}
      \caption{The same as in Fig.~\ref{fig:time_evolutions}, but for the
        DD2 EOS.}
 \label{fig:time_dd2}
\end{figure}

To further quantify these observations, we report in
Fig.~\ref{fig:mudelta_xy_avg} here the evolution of the mass-averaged
potentials in the $(x,y)$-plane, $\langle\mu_{\Delta}\rangle_{\rm xy}$,
which describe the $npe\nu$ and $np\mu\nu$ equilibria in both the $5$-$\nu$
and $3$-$\nu$ scenarios for the SFHo (top panel) and DD2 (bottom panel) EOSs. 
Despite residual contributions from regions with $\rho <
10^{14}\,\rm{g~cm^{-3}}$, we find that $\langle\mu_{\Delta}\rangle_{\rm
  xy}$ converges to a nearly constant value close to $0$ for both
equilibria. This occurs on a timescale of $\bar{t} \approx
7$--$9\,\rm{ms}$ ($\bar{t} \approx 11$--$17\,\rm{ms}$) for the SFHo (DD2)
EOS, which is indicated by the shaded regions. We explain the longer
timescale for DD2 to achieve equilibria in terms of a lower rest-mass
density and temperature of the matter, which results in a
reduction of weak interaction rates. More importantly, both $5$-$\nu$ and
$3$-$\nu$ scenarios achieve $npe\nu$ equilibrium on similar timescales,
affirming the robustness of these equilibrations across scenarios and
EOSs.

Interestingly, the timescales over which $\langle \mu_{\Delta}^{np\mu\nu}
\rangle_{\rm xy} \approx 0$ correlate with the decrease in the muonic
neutrino luminosities $\mathcal{L}_{\nu_\mu}$ and
$\mathcal{L}_{\bar{\nu}_\mu}$ shown in the bottom part of Fig.~\ref{fig:time_evolutions}
for the SFHo EOS and in the similar Fig.~\ref{fig:time_dd2} shown here for the DD2
EOS. These reduced luminosities follow from the fact that muonic
$\beta$-processes, primarily occurring in high-density regions, reach
equilibrium from both sides of the reaction processes for (anti)neutrino
absorption and (anti)lepton-capture. 
Overall, the stiffer DD2 EOS results in lower rest-mass
densities and weaker heating at merger, reducing $\mu^-$-$\mu^+$ pair
production, as muonic contributions in the EOS have less effect on the
specific internal energy and pressure for matter at lower density and
lower temperature. This also explains why a weaker muonization 
leads to a smaller increase in $Y_\mu$ during the merger.

The impact of this weaker muonization as a result of the stiffer DD2 EOS
can also be appreciated also from the bottom panel of Fig.~\ref{fig:time_dd2}
here, which shows that, unlike the SFHo $5$-$\nu$ scenario,
$\mathcal{L}_{\nu_\mu}$ for the DD2 $5$-$\nu$ scenario exceeds
$\mathcal{L}_{\bar{\nu}_\mu}$ slightly, and the luminosities of muon
flavor (anti)neutrinos are reduced, amounting to only about $50\%$ of
$\mathcal{L}_{\bar{\nu}_e}$ at the early stage. On the other hand,
$\mathcal{L}_{\bar{\nu}_e} \approx \mathcal{L}_{\nu_\mu} \approx
\mathcal{L}_{\bar{\nu}_\mu}$ in the first $\bar{t} = 7 \, \rm{ms}$ for
the SFHo EOS. Therefore, weaker muonic interactions result in smaller
differences in the total luminosity between the $3$-$\nu$ and $5$-$\nu$
scenarios, with the latter consistently showing higher luminosity until
$\bar{t} \approx 15\,\rm{ms}$. Two additional remarks are worth
making. First, the difference in $\mu_{\bar{\nu}_e}$ at $\rho >
10^{14}\,\rm{g~cm^{-3}}$ between these scenarios is also smaller for the
DD2 EOS. Second, the reduction in energy consumed by muonic processes
leaves more energy for $e^+$-capture, thereby mitigating the reduction in
$\mathcal{L}_{\bar{\nu}_e}$. Consequently, the reduction of protonization
by muonic processes is less pronounced.

\begin{figure*}
   \includegraphics[width=0.95\textwidth]{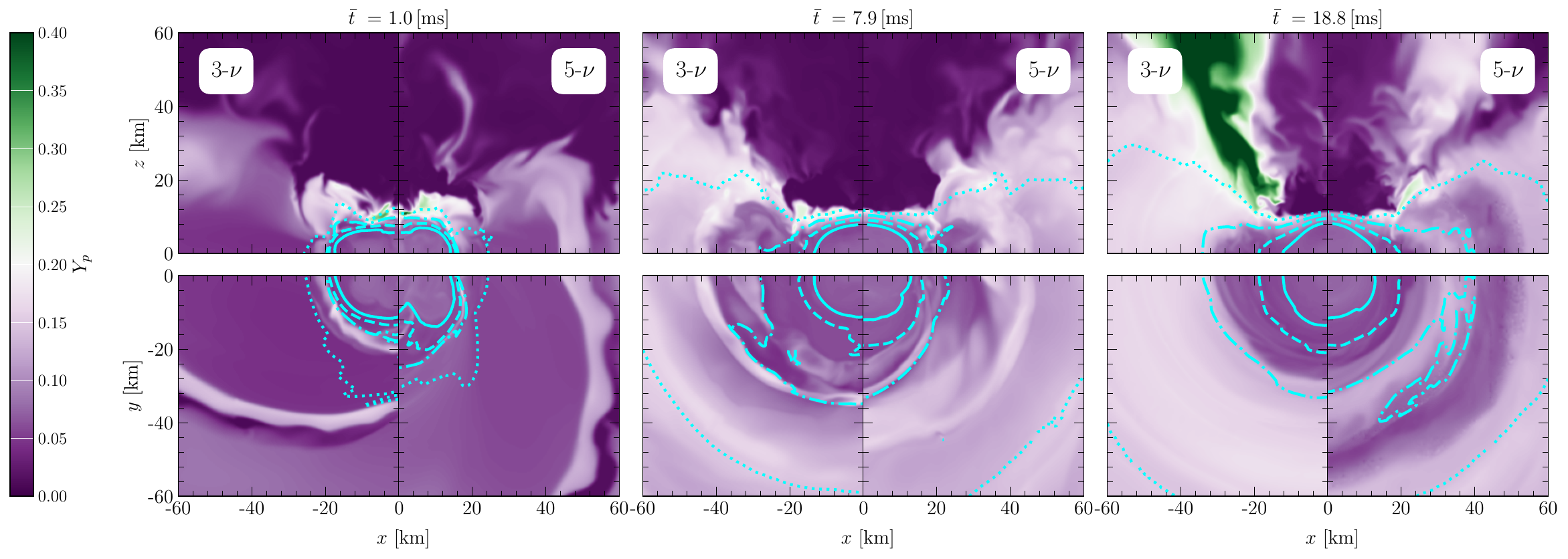}
   \caption{The same as in Fig.~\ref{fig:threemoments_sfho_yp}, but for the DD2
     EOS. Note that the five neutrino-species scenario shows a smaller
     decrease in $Y_p$ when compared to the SFHo EOS.}
   \label{fig:dd2_yp}
\end{figure*}

Figure~\ref{fig:dd2_yp}, which twins Fig.~\ref{fig:threemoments_sfho_yp}
but for the DD2 EOS, shows a behavior for the proton fraction that is
qualitatively very similar to that already discussed for the SFHo EOS,
but showing smaller differences between the $5$-$\nu$ and $3$-$\nu$
scenarios as a result of the milder densities and temperatures. Overall,
this confirms the robustness and self-consistency of our modeling of
muonic interactions both at the level of the EOS and of the neutrino
transport.

\begin{figure*}
\includegraphics[width=0.99\textwidth]{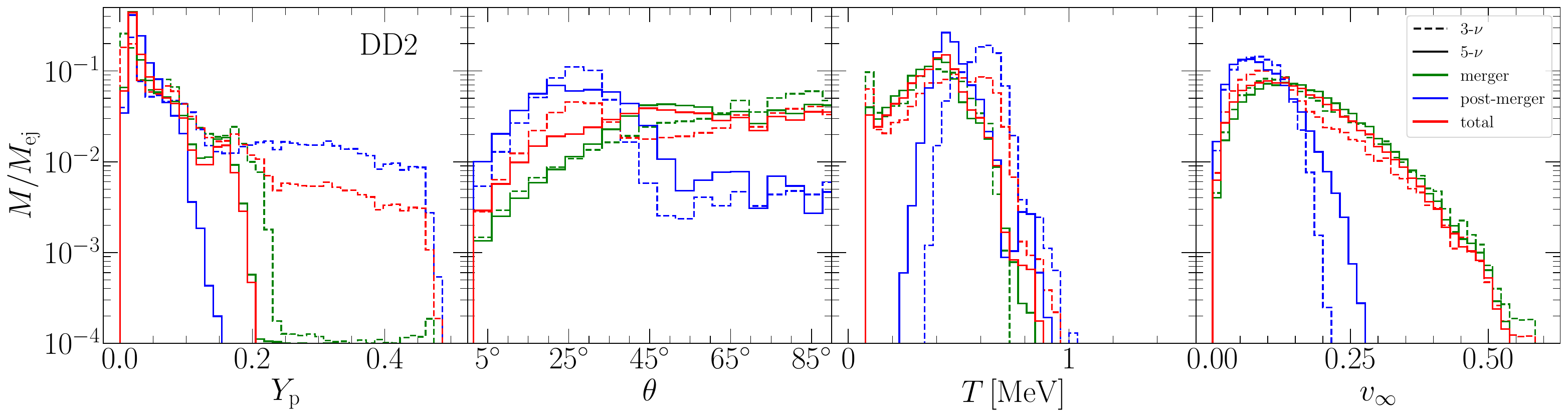}
\caption{The same as in Fig.~\ref{fig:histogram_sfho} but for the DD2
  EOS. Note that unlike SFHo cases, red lines are the total
  distributions, while green and blue lines refer respectively to
  $\bar{t} \leq 6.5\,\rm{ms}$ and $\bar{t} \geq 6.5\,\rm{ms}$. Similar to
  SFHo cases, $Y_\mu$ is negligible, and $Y_p \approx Y_e$.}
\label{fig:histogram_dd2}
\end{figure*}

Figure~\ref{fig:histogram_dd2} offers a view of the ejecta properties for
the DD2 EOS and hence represents the equivalent of
Fig.~\ref{fig:histogram_sfho} for the second EOS considered in our study.
Similar to what is seen in SFHo cases, the neutron-rich ejecta
accounts for a larger fraction of the ejecta and unbound matter when $Y_p
> 0.2$ is strongly suppressed, both before and after the merger. This
results in ejecta with a slightly lower $T$ and a less concentrated
distribution at high latitudes ($\theta < 45^{\circ}$). These
considerations demonstrate that the ejecta composition largely depends on
the differences in the merger and postmerger phenomenology and therefore
on the EOS.

\begin{figure}
\includegraphics[width=0.45\textwidth]{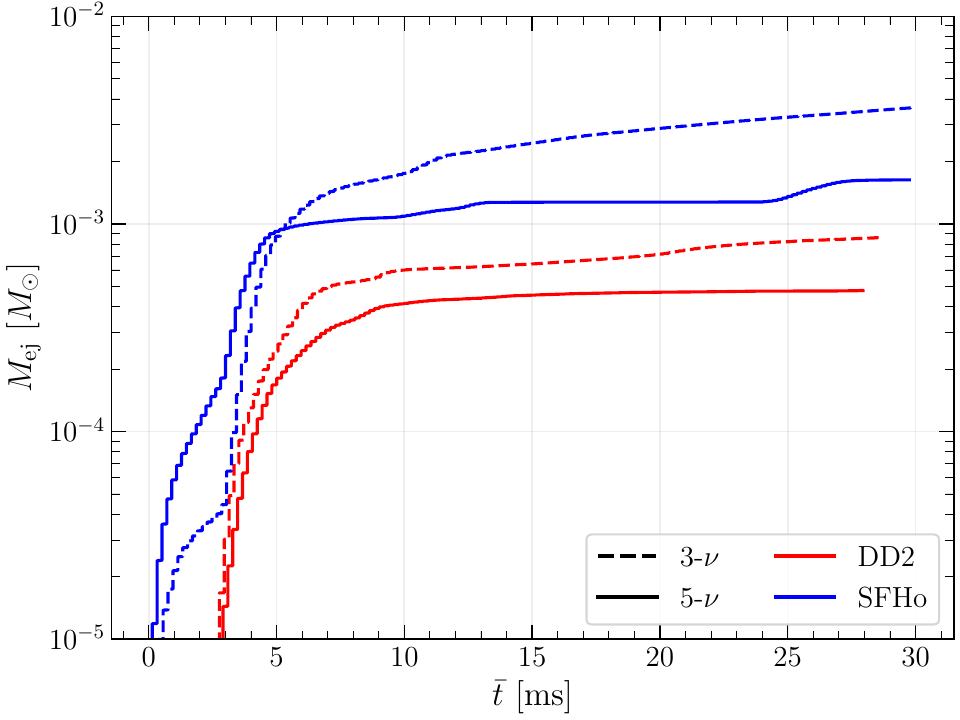}
    \caption{Comparison of the accumulated ejecta mass for $5$-$\nu$ and
      $3$-$\nu$ scenarios and for the DD2 and SFHo EOSs. Here we find
      that $5$-$\nu$ consistently results in lower postmerger ejecta for both
      EOSs.}  \label{fig:meje_ts}
\end{figure}

To complement Figs.~\ref{fig:histogram_sfho} and \ref{fig:histogram_dd2},
Figure~\ref{fig:meje_ts} shows the time evolution of the accumulated
ejecta mass for both the $3$-$\nu$ and $5$-$\nu$ simulations using the
SFHo and DD2 EOSs.

The total ejecta masses for the $5$-$\nu$ ($3$-$\nu$) scenario are, 
respectively, $1.63 \times 10^{-3}\, M_{\odot}$ ($3.64 \times 10^{-3}\,
M_{\odot}$) for the SFHo EOS and $4.8 \times 10^{-4}\, M_{\odot}$
($8.6 \times 10^{-4}\, M_{\odot}$) for the DD2 EOS. A smaller amount of
ejected mass for stiff EOSs is not surprising, especially when
considering neutrino emission, and has been reported in a number of
studies~\citep{Baiotti2016}. Quite generically, this reduction arises from
weaker shocks during the merger and postmerger, but also from remnants
that are cooler (see also~\citet{Espino2024a}). In our simulations, this
corresponds to temperatures that in the DD2 case are $\simeq
5 \, \mathrm{MeV}$ ($\simeq 1 \, \mathrm{MeV}$) lower than for the softer
SFHo EOS in the $5$-$\nu$ ($3$-$\nu$) scenario.

However, what is novel and important is the fact that, for both EOSs, the
$5$-$\nu$ simulations yield significantly less ejected matter than the
$3$-$\nu$ scenarios. Also different is the amount of matter ejected
around the merger from the matter ejected on longer timescales. In
particular, in the case of the simulations with five neutrino species and
the soft EOS, a larger amount of matter is ejected at the merger when
compared with the scenario employing three neutrino species. This
difference is not present in the case of the stiffer DD2 EOS. Additionally,
in the DD2 simulations, the postmerger ejecta (\ie reaching the detector
in the first $\sim 5\,{\rm ms}$) account for only {$39\%$ ($44\%$)} of
the total ejecta mass for the $5$-$\nu$ ($3$-$\nu$) scenario, while with the
SFHo EOS, they contribute {$51\%$ ($80\%$)} for the $5$-$\nu$ ($3$-$\nu$)
scenario. This difference is driven mostly by the muonic interactions
that emit more neutrinos in the early postmerger phase and that lead to
a cooler remnant and disk.

\begin{figure}
  \includegraphics[width=1.0\columnwidth]{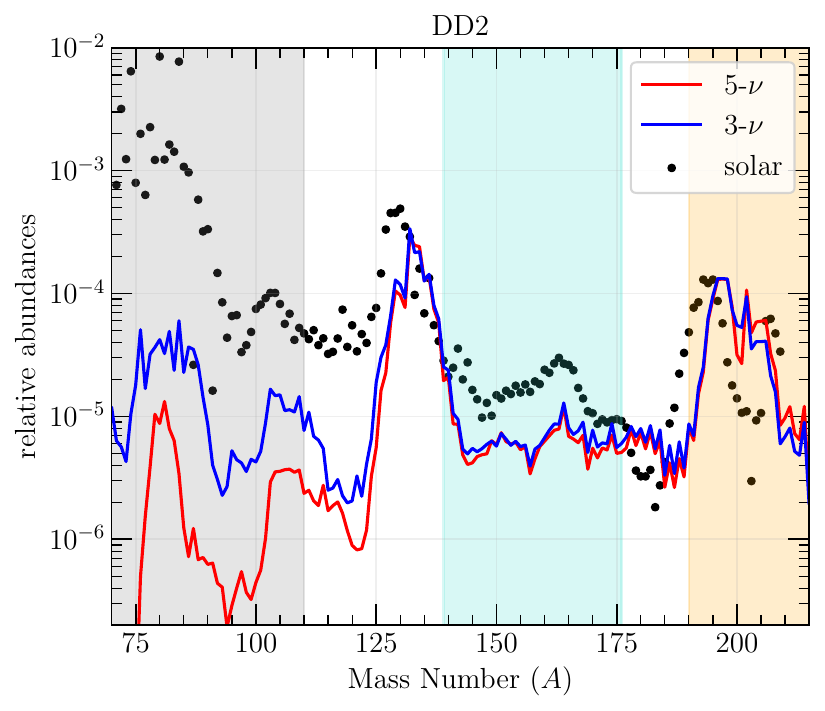}
  \caption{The same as in Fig.~\ref{fig:abundance} but for the DD2 EOS.}
\label{fig:abundance_dd2}
\end{figure}

Finally, we discuss the chemical composition of the ejected matter after
it has undergone $r$-process nucleosynthesis. This is shown in
Fig.~\ref{fig:abundance_dd2}, which twins Fig.~\ref{fig:abundance} and
that is relative to the SFHo EOS. We have already mentioned in the main
text that in the latter case, the inclusion of five neutrino species
leads to larger abundances of elements with mass number $139 < A < 176$
(blue shaded region), and to almost an order-of-magnitude larger yields
for elements with $190 < A < 215$ (orange shaded region). However, in the
case of the stiffer DD2 EOS, these differences are much smaller, and,
indeed, we measure lanthanide yields that are reduced by $\sim 10\%$ when
considering five neutrino species.

This different behavior can be explained in a number of ways: first, via
the lower temperatures and densities producing weaker shocks for the DD2
EOS (see also~\citet{Espino2024a}), which eject significantly less
matter; second, via the smaller impact of muonization, that intrinsically
requires higher temperatures and densities, hence is more prominent in
softer EOSs; and finally, via the smaller reduction of the proton fraction in the
postmerger that naturally impacts nucleosynthesis. As
a concluding remark, we note that the lack of a ``lanthanide boost'' in
the stiffer DD2 EOS may also be the result of the limited timescale over
which our simulations have been carried out. Indeed, it is possible that
muonic interactions may require longer timescales to increase lanthanide
production in stiff EOSs. We plan to explore this conjecture in future
works covering longer postmerger timescales.

\section{Strategies to prevent over-(de)muonization}
\label{sec:strategy}

When $\mu_e \le m_\mu$, initial data in both $np\mu$ and $npe$ equilibria
must abandon the $np\mu$-equilibrium condition by setting $Y_\mu = 0$ to
asymptotically satisfy $npe$-equilibrium, thus matching composition in
low-density, low-temperature regions without (anti)muons. This adjustment
is necessary because both equilibria cannot coexist in these regions:
$\mu_e$ can never equal $\mu_\mu$ when $Y_\mu$ is nonzero (see Fig.~3.5
in \citet{Bolligthesis}). During the evolution, unphysically high (or
low) values of $\eta^{\rm eq}_{\nu_\mu}$ (or $\eta^{\rm
  eq}_{\bar{\nu}_\mu}$)~\citep{Loffredo2023}, where $\eta^{\rm
  eq}_{\nu_\mu} := (\mu_\mu + \mu_p - \mu_n)/T$ and $\eta^{\rm
  eq}_{\bar{\nu}_\mu} = - \eta^{\rm eq}_{\nu_\mu}$, arise in low-density,
low-temperature regions, where $\mu_\mu \sim m_\mu$ and $Y_\mu \sim
0$. Under these conditions, and especially at the surfaces of stars
during the inspiral, $\eta^{\rm eq}_{\nu_\mu}$ can reach values $\sim
5-100$, causing excess $\mu^-$ production with $Y_\mu$ hitting the upper
bound of the table. As argued by~\citet{Bolligthesis}, tracking vanishing
$Y_\mu$ with weak interactions for these regions is unfeasible, and
suitable strategies have to be designed and implemented.

In our approach, we first define a threshold $Y^{\rm thr}_\mu$ and set
$\eta^{\rm eq}_{\nu_\mu} = \eta^{\rm eq}_{\bar{\nu}_\mu}= 0$ in those
regions where $Y_\mu < Y^{\rm thr}_\mu$; experimentation has shown that
setting $Y^{\rm thr}_\mu = 5\times 10^{-3}$ avoids excessive increases in
the muon fraction. Furthermore, following Eq.~(9.31) in
~\citet{Bolligthesis}, we apply a suppression factor based on the
rest-mass density and temperature cutoffs to muonic weak interactions of
the type $\mathcal{Q}^{\prime} = \mathcal{Q} [1/(1 + (\rho_{\rm
    th}/\rho)^5] [1/(1 + (T_{\rm th}/T)^6]$, where $\mathcal{Q}$ refers
to quantities such as the energy-averaged emissivities $\bar{Q}$, or the
energy-averaged absorption opacities $\bar{\kappa}_{a}$, and where we set
$\rho_{\rm th} := 10^{11}\, \mathrm{g}~\mathrm{cm}^{-3}$ and $T_{\rm th}
:= 2.5\,\mathrm{MeV}$.

Another issue to consider is that energy-averaged (or gray) moment
schemes such as the one employed here depend on degeneracy parameters for
detailed balance (Kirchhoff's law), the energy averaging of the
opacities, and the rate corrections. Within three-species approaches, it
is not necessary to rescale the degeneracy parameters for $\nu_e$ and
$\bar{\nu}_e$, as the small electron mass leads to a monotonic decrease
in $\eta^{\rm eq}_{\nu_e}$ as the rest-mass density reaches the
atmosphere values. However, in five-species approaches, very large values of
$\eta^{\rm eq}_{\nu_\mu}$ and $\eta^{\rm eq}_{\bar{\nu}_\mu}$ need
corrections in regions where muonic interactions are negligible and
$np\mu$ equilibrium is absent. Thus, we rescale the degeneracy parameters
$\eta_{\nu}$ using the optical depth $\tau_\nu$, which is computed by
approximately solving iteratively the Eikonal equation of $\nabla
\tau_{\nu}(x) = \bar{\kappa}_{\nu}(x)$~\citep{Neilsen2014} by the method
of~\citet{Zhao2005}, where $\bar{\kappa}_{\nu}(x)$ is the total
energy-averaged opacity as a function of spatial coordinate $x$.
$\eta_\nu$ is thus expressed as $\eta_\nu = \eta^{\rm eq}_\nu ( 1
-e^{-\tau_{\nu}})$ for all $\nu$, and we set $\eta_{i} = 0$ where $\tau_i
< 1$, and $i$ only refers to $\nu_\mu$ or $\bar{\nu}_\mu$.

Finally, due to the missing spectral information about the distribution
function of neutrinos, attention needs to be paid with Kirchhoff's law
for pair processes (e)-(g) in Table.~\ref{tab:weak_interactions}.
A possibility is to use an approximated isotropic
emissivity~\citep{Oconnor2015} instead of the inelastic kernels for pair
processes. On the other hand, \texttt{Weakhub}~\citep{Ng2024a} or
\texttt{NuLib}~\citep{Oconnor2015} calculates the pair process spectral
absorption opacity $\kappa_{a}^{\rm{pp}}$ by first computing the spectral
emissivity $Q^{\rm pp}$ and then applying Kirchhoff's law
$\kappa_{a}^{\rm{pp}} = Q^{\rm pp} / \mathcal{B}(\eta^{\rm eq}_\nu)$,
where $\mathcal{B}(\eta^{\rm eq}_\nu)$ is the blackbody spectrum. This
quantity is tabulated as an energy-averaged opacity and stored similarly
to other absorption opacities. However, in low-density regions, values of
$-100 \lesssim \eta^{\rm eq}_{\bar{\nu}_\mu} \lesssim -5$ yield
$\mathcal{B}(\eta^{\rm eq}_{\bar{\nu}_\mu}) \simeq 0$ and hence
excessively high values of $\kappa_{a, \bar{\nu}_\mu}^{\rm{pp}}$, causing
extreme demuonization and destroying $np\mu\nu$-equilibrium after the
merger. To counter this, we calculate $\kappa_{a}^{\rm{pp}}$ on the fly
with the fixed $\eta_\nu$ instead of $\eta^{\rm eq}_\nu$ (alternatively,
the isotropic emissivity could be tabulated and the absorption opacity be
recalculated on the fly with $\eta_\nu$).

\bibliography{aeireferences}

\end{document}